\documentclass[12pt,preprint]{aastex}

\shorttitle{Formation of $\omega$ Centauri}
\shortauthors{Ideta \& Makino}

\begin{document}

\title{Formation of $\omega$ Centauri by Tidal Stripping of a Dwarf
Galaxy}
\author{Makoto Ideta}
\affil{Astronomical Data Analysis Center, National Astronomical
Observatory of Japan\\2-21-1 Osawa, Mitaka, Tokyo 181-8588, Japan}
\and
\author{Junichiro Makino}
\affil{Department of Astronomy, University of Tokyo\\7-3-1 Hongo,
Bunkyo-ku, Tokyo 113-0033, Japan}

\begin{abstract}
 We have investigated whether or not a tidal stripping scenario can
 reproduce the observed surface-brightness profile of $\omega$ Centauri
 using $N$-body simulations.  Assuming that the progenitor of $\omega$
 Centauri is a dwarf elliptical galaxy, we model it with a King model
 with a core radius being the same as that of $\omega$ Centauri.  A dark
 matter halo of the dwarf is not taken into account.  We consider two
 different models of the Milky Way potential: a singular isothermal
 sphere and a three-component model.  The progenitor dwarf is expressed
 as an $N$-body system, which orbits in the fixed Galactic potential.
 The dwarf lost more than 90 per cent of its mass during the first few
 pericenter passages.  Thereafter, the mass remains practically
 constant.  The final surface-density profile is in good agreement with
 the observational data on $\omega$ Centauri, if the pericenter distance
 of the orbit of the progenitor dwarf is around $500\;{\rm pc}$.  This
 value is within the error bar of the current proper motion data on
 $\omega$ Centauri and Galactic parameters.  Although our simulation is
 limited to a King-like progenitor dwarf without a dark matter halo, it
 strongly suggests that the current density profile of $\omega$ Centauri
 is nicely reproduced by a tidal stripping scenario, in other words,
 that $\omega$ Centauri can plausibly be identified with a stripped
 dwarf elliptical.
\end{abstract}


\section{INTRODUCTION}
$\omega$ Centauri (NGC 5139) is, to the best of our knowledge, the most
massive globular cluster in the Milky Way, with a mass around
$5\times10^6M_\odot$ \citep{meylan95}.  Furthermore, it shows peculiar
chemical and dynamical features, such as the wide spread in metallicity
distribution of member stars \citep{freeman81} and the difference in
kinematics of metal-rich and metal-poor populations \citep{norris97}.
Some member stars show an enhancement of s-process elements produced by
AGB stars \citep[e.g.,][]{smith00}, which means that the potential well
must have been deep enough to trap the gas ejected from AGB stars
\citep{gnedin02}.  A merger model \citep{icke88} can explain the
kinematics and metallicity spread, but has difficulty in accounting for
the presence of s-process elements.

A tidal stripping scenario \citep{zinnecker88,freeman93} has been
proposed to explain all observed features simultaneously.  In this
scenario, the cluster was born as a dwarf elliptical galaxy, which was
much more massive than the current $\omega$ Centauri.  As it sinks to
the Galactic center through dynamical friction, it loses most of its
mass through the tidal stripping by the gravitational potential of the
Milky Way.  However, if the central density is sufficiently high, the
core of the dwarf can still survive as a bound system and become a
relatively large globular cluster, something like $\omega$ Centauri.  In
this scenario, a relatively long star formation history is naturally
explained because the potential well of the progenitor dwarf elliptical
was initially much deeper than that of the present-day $\omega$
Centauri.  In addition, since the progenitor dwarf could be formed
through hierarchical merging events, signatures of past merging events
can also be explained.  Furthermore, recently, \citet{martini04}
reported that some of massive globular clusters in NGC 5128 may also be
such tidally stripped dwarf ellipticals.

From the viewpoint of dynamics, whether or not the stripping scenario
works depends on the following two questions.  The first one is whether
the dynamical friction can bring the progenitor dwarf to the current
orbit of $\omega$ Centauri.  The second one is whether the tidal
stripping can actually produce the observed surface brightness of
$\omega$ Centauri.  The first question has been addressed by recent work
\citep*{zhao02,tsuchiya03,bekki03}.  However, the second question has
not been studied so far, despite its significance to the validity of the
stripping scenario.

In this paper, we have investigated whether a dwarf galaxy with the
orbit similar to that of the present-day $\omega$ Centauri will evolve
to have the spatial structure which agrees with that of $\omega$
Centauri using $N$-body simulations.  In \S~2, numerical models and
method are described.  Results are shown in \S~3.  A discussion is given
in \S~4.

\section{MODELS AND METHOD}
We consider the dynamical evolution of a dwarf elliptical in a fixed
potential.  In this paper, we neglect the effects of dynamical friction.
This neglect should not affect results as far as a progenitor dwarf
along the present-day orbit of $\omega$ Centauri is concerned, since our
main goal is to see whether or not the tidal field of the Milky Way in
the present-day orbit of $\omega$ Centauri can account for its density
structure.  In reality, dynamical friction must have played a certain
role in bringing the progenitor dwarf to its current orbit, but that
part of evolution is not in the scope of the present paper.

We use the data compiled by \citet{meylan87} for the surface-brightness
profile of $\omega$ Centauri.  This profile is well-fitted (except in
the outermost region) by a King model with a non-dimensional central
potential $W_0=5.5$ and a core radius $R_c=4.6\;{\rm pc}$.  We choose
the total mass of this King model to be identical to the mass suggested
from observations, $5.0\times10^6\;M_{\odot}$.  Then, we set up the
model of the progenitor dwarf to have the same core radius and central
velocity dispersion, but with a much more extended halo.  We choose a
King model with $W_c=12$ as a model for the dwarf.  Since some dwarf
elliptical galaxies can also be fitted by a King model, we believe this
choice is justified.  \citet*{tsuchiya04} have found that a King-like
dwarf elliptical cannot evolve to $\omega$ Centauri on the basis of
numerical simulations of King- and Hernquist-like dwarf galaxies
embedded in a live Milky Way potential.  This point will be discussed in
\S~4.1.

The dwarf elliptical is represented by $2^{20}\simeq1\;{\rm million}$
particles of equal mass.  The initial total mass is
$1.3\times10^8\;M_{\odot}$.  We do not include a dark matter halo of the
dwarf, which would be more extended than the distribution of visible
stars.  Since the extended halo would be stripped out anyway, we believe
the presence of a dark matter halo does not change results
significantly.  In most of our simulations, the final cluster still had
more than 50,000 particles, with the result that the relaxation time is
much longer than the duration of the simulation.

\begin{figure}
 \plotone{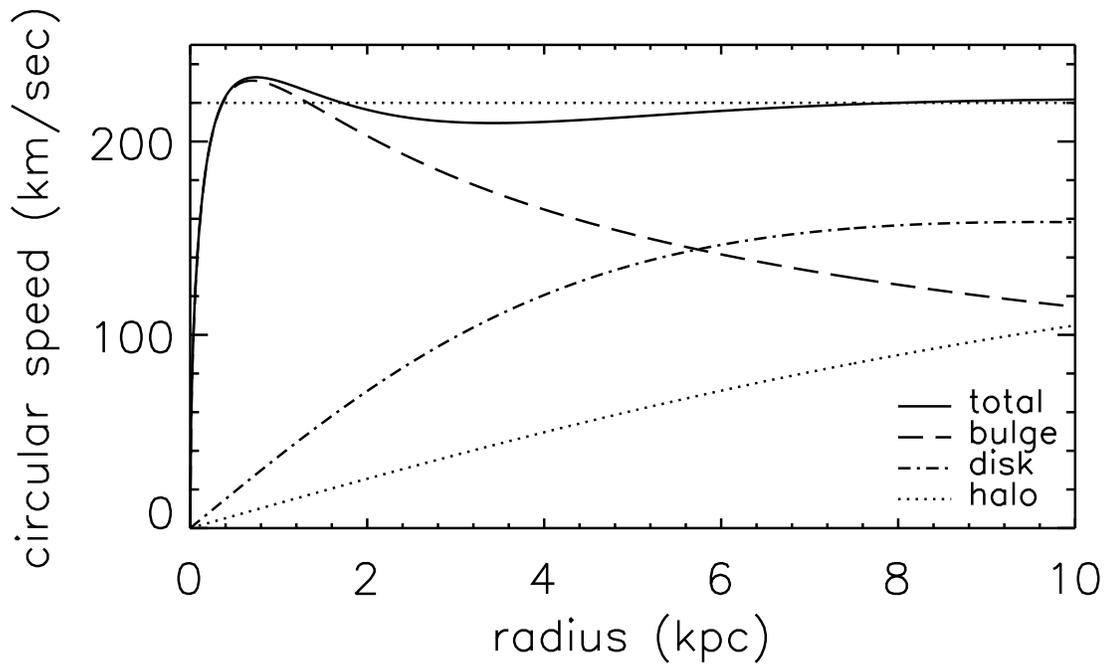}
 \figcaption[f1.eps]{Circular speed curves of our Galactic models.  The
 dashed, dash-dotted, and dotted curves give the contributions of the
 bulge, disk, and halo, respectively, while the solid curve gives the
 total circular speed in the three-component model.  The horizontal
 dotted line at $220\;{\rm km\;sec^{-1}}$ is the rotation curve of the
 isothermal model.  \label{fig:model}}
\end{figure}

We use two different models for the gravitational potential of the Milky
Way, a singular isothermal sphere (run std) and a bulge-disk-halo model
(run mwd).  The circular velocity of the std potential is $220\;{\rm
km\;sec^{-1}}$.  The mwd potential is almost the same as the model used
in \citet*{jsh95}.  This model consists of three components: a
Miyamoto-Nagai \citeyearpar{mn75} disk, a Hernquist \citeyearpar{lh90}
bulge, and a logarithmic halo potential.  The only difference between
the model of \citet{jsh95} and ours is the core radius of the
logarithmic potential.  We set the core radius to be $14.0\;{\rm kpc}$,
while \citet{jsh95} set it to be $12.0\;{\rm kpc}$, so that the circular
velocity at the solar circle radius, i.e., $8.0\;{\rm kpc}$, is
$220\;{\rm km\;sec^{-1}}$.  The potentials and parameters used in
simulations are listed in Table~\ref{tab:model}.  The circular speed
curves of these models are plotted in Figure~\ref{fig:model}.

We use the following kinematical data on $\omega$ Centauri: the distance
from the Sun is $4.9\;{\rm kpc}$, the proper motions are
$(\mu_\alpha\cos{\delta},\mu_\delta)=(-5.08\pm0.35,-3.57\pm0.34) \;{\rm
mas \; yr^{-1}}$, and the radial velocity is $232.2\pm0.7\;{\rm
km\;sec^{-1}}$ \citep{dinescu99}.  For the most probable values of the
proper motions and the radial velocity in the isothermal model (run
std), the pericenter and apocenter distances are $1.0\;{\rm kpc}$ and
$6.4\;{\rm kpc}$, respectively.  We made two other runs, in which we
expand the initial dwarf by a factor of 1.5 (run r15) and 2 (run r20),
to see the effect of the change in the tidal force at the pericenter.
For the singular isothermal model, the only characteristic scale in the
experiment is the core radius of the cluster.  A linear expansion of the
cluster by a factor of 1.5 or 2 can therefore equally well correspond to
a reduction in the pericenter distance of the unmodified cluster by a
factor of $1.5^{3/2}\simeq2$ (run r15) or $2^{3/2}\simeq3$ (run r20),
which corresponds to $\sim600\;{\rm pc}$ and $\sim400\;{\rm pc}$,
respectively.

We use a hierarchical tree algorithm \citep{bh86} on the GRAPE-6
hardware \citep{makino03} with an opening angle being 0.5.  We employ a
Plummer softening of $0.5 \;{\rm pc}$, which is roughly one-tenth of the
observed core radius of $\omega$ Centauri, $4.6\;{\rm pc}$.  The
equations of motion are integrated using a leap-frog method with a
constant time-step of $\Delta t=1.7\times10^4\;{\rm yr}$.  This
time-step is comparable to the time required for a particle with the
maximum circular velocity around $\omega$ Centauri to cover the
softening length.  The total energy was conserved to better than $0.005$
per cent in all simulations.

\section{RESULTS}
\begin{figure}
 \plotone{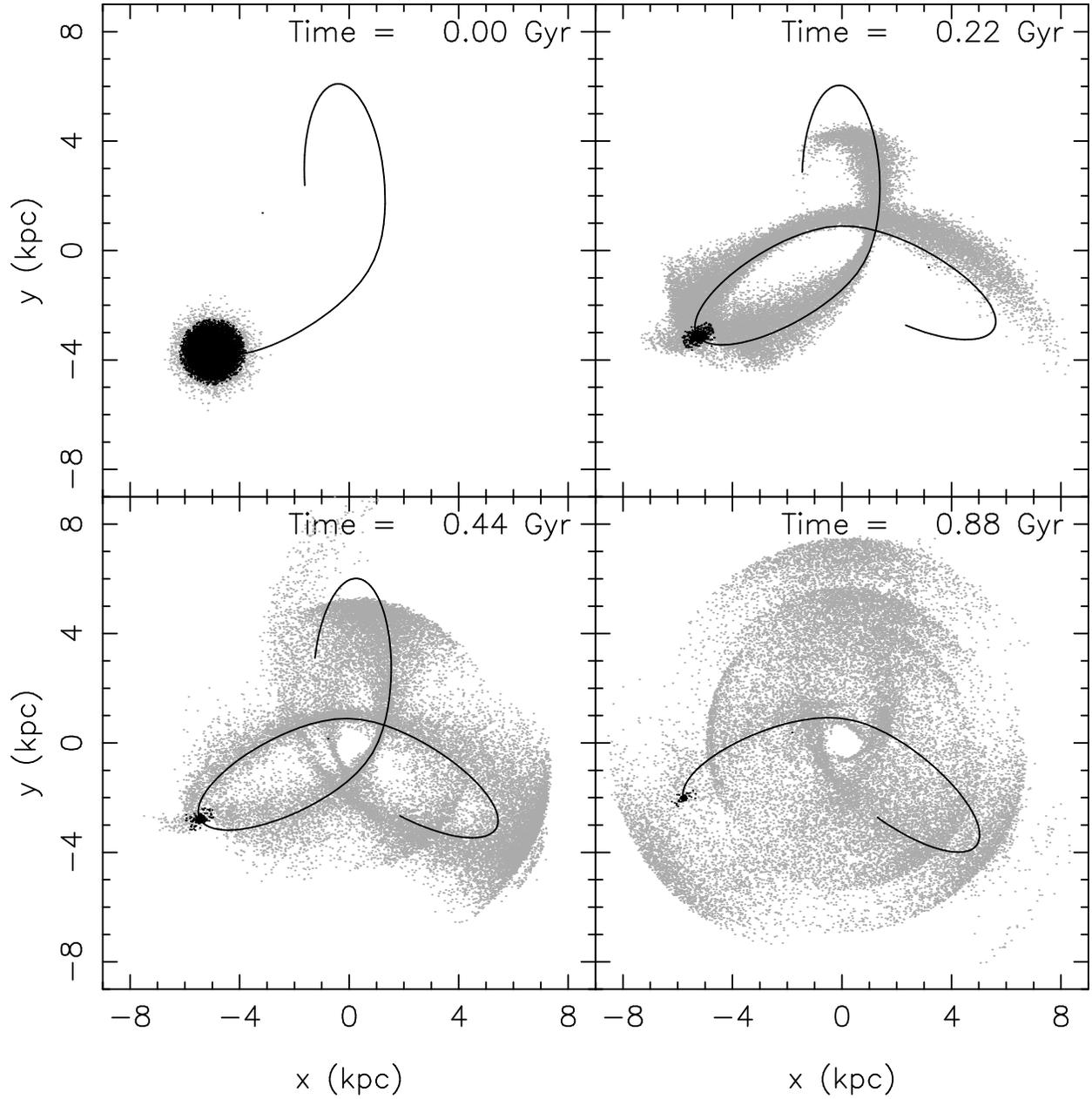}
 \figcaption[f2.eps]{Snapshot images from run std.  Bound particles are
 drawn in black, while unbound ones are in gray.  Simulation time is
 shown at the top-right corner of each box.  The orbit of the cluster
 center for periods of 0.1 Gyr before and after each simulation time is
 also drawn in the solid line.  \label{fig:snap}}
\end{figure}

Figure~\ref{fig:snap} shows snapshot images from run std, along with the
orbit of the cluster center for the periods of 0.1 Gyr before and after
the time for the snapshot (whenever the orbit is available).  The
cluster center is defined as the position of the particle with the
minimum potential energy for the $N$-body particles.  We include all
particles to calculate the potential.  Removing unbound particles did
not affect the result, even after more than 90 per cent of the particles
were stripped.  We can see that the stripped stars remain close to the
cluster orbit, and form numerous ripple structures.

\begin{figure}
 \plotone{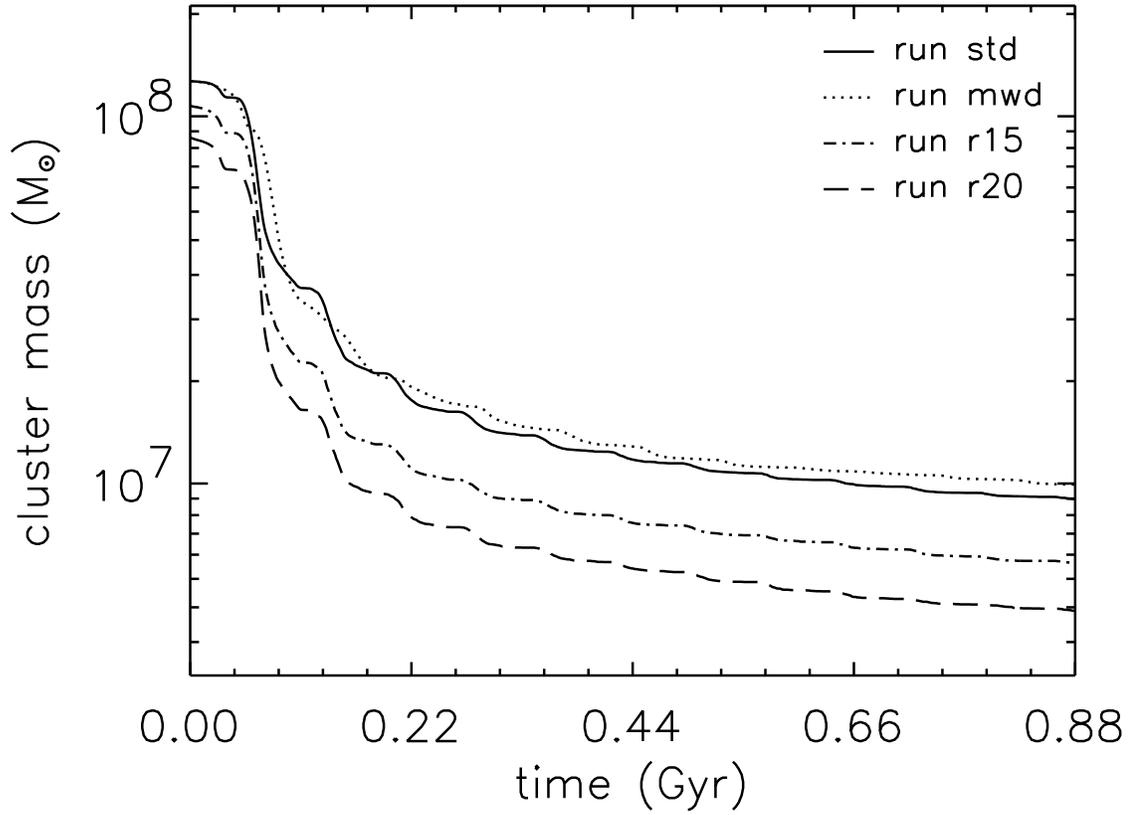}
 \figcaption[f3.eps]{Time evolution of the cluster mass for all runs.
 The solid, dotted, dash-dotted and dashed curves represent the results
 of runs std, mwd, r15, and r20, respectively.  \label{fig:mass}}
\end{figure}

Figure~\ref{fig:mass} shows the time evolution of the cluster mass.
Here, we define the cluster mass simply as a mass within $2 r_t$, where
$r_t$ is the tidal radius at the apocentric position expressed as
\begin{equation}
 r_t = \left(\frac{m_{\rm c}}{2M_{\rm G}}\right)^{1/3}R_{\rm G}.
\label{eq:rt}
\end{equation}
Here $R_{\rm G}$ is the distance from the Galactic center to the
cluster, $M_{\rm G}$ is the Galaxy mass within $R_{\rm G}$, and $m_{\rm
c}$ is the cluster mass.  We assume that the stars are stripped from the
cluster if their distances from the cluster center are larger than two
apocentric tidal radii \citep[e.g.,][]{holger03a}.  For the
three-component model, we have determined the cluster mass in the same
way.  The cluster mass exhibits periodic sudden drops, which correspond
to the pericentric passage.  Thus, almost all the mass loss occurred at
the pericentric passage.  Since the cluster is expanded in runs r15 and
r20, the mass loss is larger for these runs than that for run std.
There is very little difference between the result of run std and that
of mwd, suggesting that the disk does not contribute significantly to
the tidal force.  This result is of course not surprising, because the
disk gravity is small at a pericenter distance of $1\;{\rm kpc}$.

\begin{figure}
 \plotone{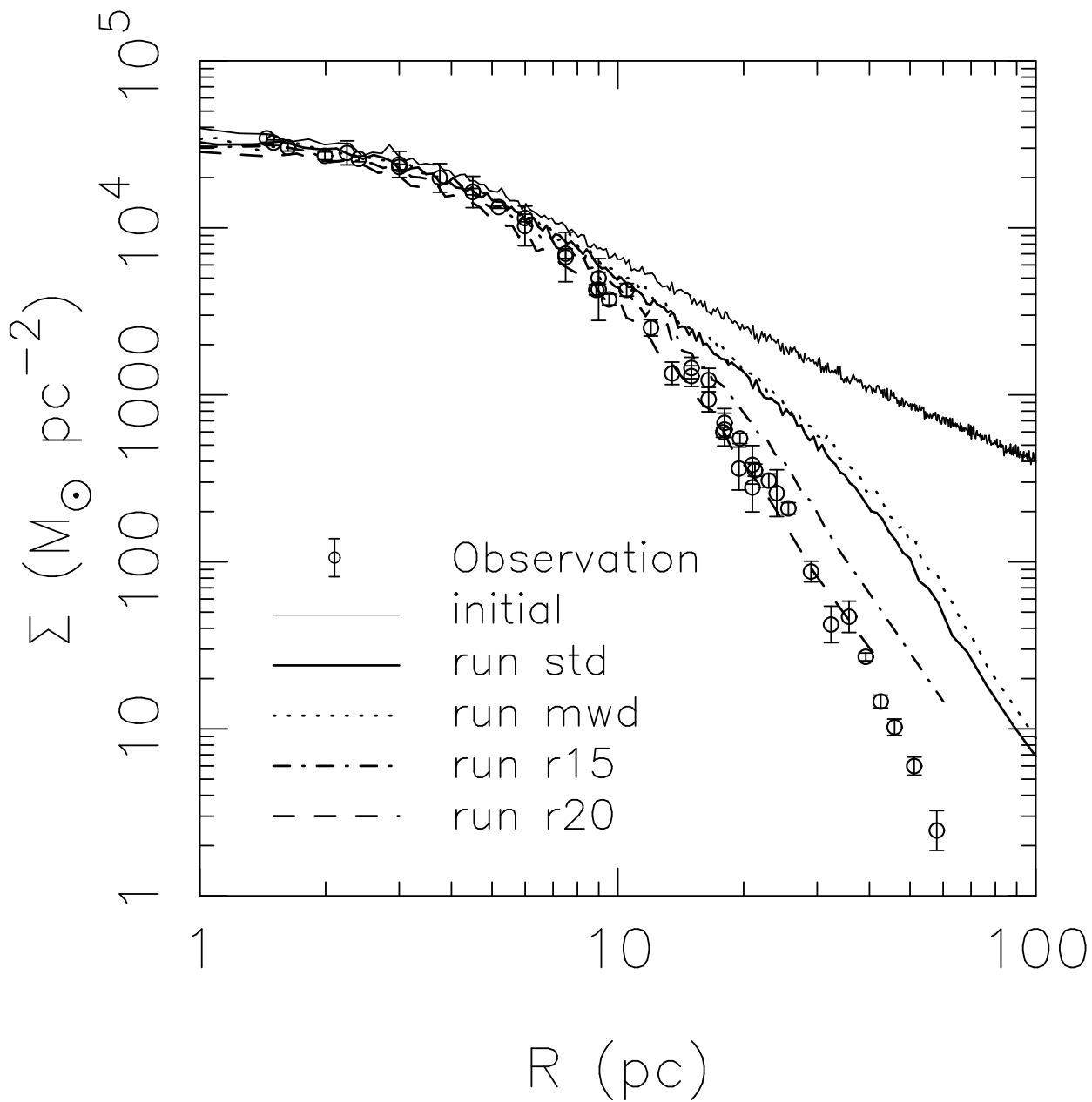}
 \figcaption[f4.eps]{Final surface-density (brightness) profiles of the
 clusters for all runs.  The curves have the same meanings as in figure
 3.  The open circles with error bars show the surface-density profile
 calculated from the observed surface brightness on the assumption of a
 constant mass-to-light ratio.  The thin solid line corresponds to the
 initial model.  \label{fig:sden}}
\end{figure}

Figure~\ref{fig:sden} shows the final surface-density profiles for all
runs.  We assume a constant mass-to-light ratio, since the estimated
relaxation time of $\omega$ Centauri is too long for the mass
segregation to be visible \citep[e.g.,][]{gh03}.  Clearly, the result of
run std does not agree with observations.  Replacing the Galactic
potential with a more realistic one (run mwd) does not improve the
situation.  On the other hand, changing the pericenter distance has a
drastic effect, and the agreement of run r20 with observations is very
good.  The result of r15 falls between std and r20.

We conclude that a tidal stripping scenario involving a King-like
progenitor without a dark matter halo can reproduce the observed
luminosity profile quite well, provided that the pericentric distance of
$\omega$ Centauri is around $400\;{\rm pc}$.

\section{DISCUSSION}

\subsection{Comparison with Previous Work}
Our conclusion that a King-like dwarf can evolve to $\omega$ Centauri
apparently contradicts with the result by \citet{tsuchiya04} that final
clusters started from King-like dwarf ellipticals are still more massive
than $10^8\;M_{\odot}$.  This difference is mainly due to the high
velocity dispersion of the initial dwarf models in their simulations.
The central velocity dispersion in their model K5 reaches $\sim55\;{\rm
km\;sec^{-1}}$, while we set it to be $22\;{\rm km\;sec^{-1}}$.  This
means that the cluster mass inside the tidal radius in their model K5 is
$2.5^3\simeq16$ times larger than the mass in our model, even if the
pericenter distance is assumed to be equal.  In terms of the fraction of
total mass lost, their results are actually quite consistent with ours.

\subsection{Pericenter Distance}
We have shown that the observed surface-brightness profile of $\omega$
Centauri is nicely reproduced by the tidal stripping scenario if the
pericenter distance of the cluster is roughly $400\;{\rm pc}$.  Here, we
discuss whether such a small pericenter distance is compatible with
observations.

We have adopted the following error estimates for observational data.
For the errors in the proper motion and the radial velocity, we have
used the data by \citet{dinescu99}: $(U, V, W) =
(-64\pm11,\,-254\pm9,\,4\pm10)\;{\rm km\;sec^{-1}}$.  Here, $(U, V, W)$
is the relative velocity of $\omega$ Centauri to the local standard of
rest.  The positive directions of the $U$, $V$, and $W$ components are,
respectively, outward from the Galactic center, toward the Galactic
rotation, and toward the Galactic north pole.  For the Galactic
parameters, \citet{reid99} estimated the ratio of the circular speed to
the solar circle radius, $\Theta_0/R_0$, on the basis of the proper
motion study of Sagittarius A$^*$ by using VLBA, and concluded that the
circular speed is $219\pm20\;{\rm km\;sec^{-1}}$, assuming that the
solar circle radius is $R_0=8.0\;{\rm kpc}$.  For the error in the
circular speed at the solar circle, we have adopted the value of
$220\pm20\;{\rm km\;sec^{-1}}$.  Since the solar circle radius would
have an error comparable to that in the circular speed, we have used
$8.0\pm1.0\;{\rm kpc}$ for the solar circle radius.

We calculated the variations of the pericentric distance due to errors
in the orbital velocity of $\omega$ Centauri, the solar circle radius,
and the circular speed at the solar circle.  The results are summarized
in Table~\ref{tab:err}.  The error in the circular speed has a strong
effect on the pericenter distance, since the orbit of $\omega$ Centauri
is highly eccentric.  A 10 per cent variation in the circular speed can
lead to 50 per cent change in the pericenter distance.  The error in the
$V$ component also has such large magnification.  Thus, our result that
the pericenter distance must be around $400\;{\rm pc}$ is compatible
with observations.

We set the total mass of $\omega$ Centauri to
$5.0\times10^6\;M_{\odot}$, following the estimate by \citet{meylan95}.
However, since this mass estimate is based on a multi-mass King model,
the total mass is probably overestimated \citep[see][]{holger03b}.
Indeed, in our best-fit King model, the central velocity dispersion
reaches $22\;{\rm km\;sec^{-1}}$, while the observed value is around
$17\;{\rm km\;sec^{-1}}$ \citep*{merritt97}.  For the observed value of
the central velocity dispersion, the total mass is around
$3\times10^6\;M_{\odot}$.  This change of the cluster mass reduces the
tidal radius by about 15 per cent ($r_t\propto m_{\rm c}^{1/3}$).  In
other words, to yield the observed tidal radius with a reduced cluster
mass, the periastron distance in the model would have to be increased by
30 per cent [from equation (\ref{eq:rt}), $R_{\rm G}\propto m_{\rm
c}^{-1/2}$].  Therefore, if we use this revised estimate of the mass of
$\omega$ Centauri, the most likely value of the pericenter distance is
around $500\;{\rm pc}$.

\subsection{Long-Term Orbit Evolution in the Three-Component Model}
In section 3, we have found that the quantitative result remains
unchanged when we change the Galactic potential model from the singular
isothermal sphere to the more realistic three-component model.  However,
since we integrated our model only for 0.88 Gyr, we could have missed
some important contributions of the non-sphericity of the potential.

We integrated the orbit of a point-mass cluster in the three-component
potential for $\sim10^{13.5}\;{\rm yr}$.  Here, such a long integration
time is required since chaotic diffusion time could be much longer than
a Hubble time.  We have found that the minimum pericentric radius is
$\simeq0.9\;{\rm kpc}$, although the orbit is chaotic.  Consequently,
the effect of chaotic diffusion is negligible.

\subsection{Globular Clusters as Probes of Galactic Structure}
We have found that the structure of $\omega$ Centauri is nicely
explained by the tidal stripping scenario, for the orbital parameters
within the observational errors.  The pericenter distance is the most
important parameter that determines the structure of the simulated
$\omega$ Centauri.  In other words, we can put a fairly tight constraint
on the pericenter distance of $\omega$ Centauri from its internal
structure.  This constraint can be converted to constraints on the
Galactic potential, such as the circular speed, if we have high-accuracy
data for the proper and radial motions of $\omega$ Centauri.

We can probably apply a similar technique to other tidally-limited
globular clusters, though in this case the modeling is somewhat more
complex since we cannot ignore thermal relaxation.  Clusters in highly
eccentric orbits are most useful, since a small change in the Galactic
parameters results in a large change in the pericenter distance.  Thus,
the dynamical simulation of the evolution of globular clusters in a
Galactic tidal field, combined with high-accuracy proper motion data
which will be available via next-generation astrometry projects, will
provide us with a unique tool to probe the structure of the Milky Way.

We thank the anonymous referee for useful comments that helped us to
improve the clarity of the paper.  We are grateful to Prof. S. Hozumi
for his critical reading of the manuscript.  All simulations were run at
the University of Tokyo.  Some part of data analysis was made on
workstations at Astronomical Data Analysis Center, National Astronomical
Observatory of Japan (ADAC/NAOJ).

\begin{deluxetable}{cc}
 \tablecaption{The Milky Way models. \label{tab:model}}
 \tablewidth{0pt}
 \tablehead{
 \colhead{isothermal sphere} & \colhead{three component}
 }
 \startdata
 $\Phi=v_c^2\ln{r}$ & 
 $\Phi_{\rm d}=-\frac{GM_{\rm d}}
 {\sqrt{R^2+\left(a+\sqrt{z^2+b^2}\right)^2}}$ \\
 $v_c=220\;{\rm km\;sec^{-1}}$ &
 $M_{\rm d}=1.0\times10^{11}M_\odot, \quad 
 a=6.5\;{\rm kpc}, \quad b=0.26\;{\rm kpc}$\\ \\
 & $\Phi_{\rm b}=-\frac{GM_{\rm b}}{r+c}$ \\
 & $M_{\rm b}=3.4\times10^{10}M_\odot, \quad c=0.7\;{\rm kpc}$ \\ \\
 & $\Phi_{\rm h}=v_{\rm h}^2\ln{\left(r^2+d^2\right)}$ \\
 & $v_{\rm h}=128\;{\rm km\;sec^{-1}}, \quad d=14\;{\rm kpc}$\\
 \enddata
\end{deluxetable}

\clearpage

\begin{deluxetable}{ccc}
 \tablecaption{The pericenter distance of $\omega$ Centauri in the
 isothermal sphere model. \label{tab:err}}
 \tablewidth{0pt}
 \tablehead{
 \colhead{observable} & \colhead{value} & \colhead{pericenter}
 }
 \startdata
 velocity $(U, V, W)$ &
 $(-64\pm11,\,-254\pm9,\,4\pm10)\;{\rm km\;sec^{-1}}$ &
 $1.0\pm0.3\;{\rm kpc}$ \\ \\
 solar circle radius  & $7.0\;{\rm kpc}$          & $0.9\;{\rm kpc}$ \\
                      & $9.0\;{\rm kpc}$          & $1.0\;{\rm kpc}$ \\ \\
 circular speed       & $240\;{\rm km\;sec^{-1}}$ & $0.6\;{\rm kpc}$ \\
                      & $200\;{\rm km\;sec^{-1}}$ & $1.5\;{\rm kpc}$ \\
 \enddata
\end{deluxetable}

\clearpage

\clearpage

\clearpage

\clearpage

\end{document}